\documentclass[aps,prd,twocolumn,groupedaddress,showpacs,floatfix]{revtex4}

\usepackage[dvips]{graphics}
\usepackage[dvips]{graphicx}
\usepackage{psfrag}
\usepackage{longtable}
\usepackage{amsfonts}

\begin{document}


\preprint{COLO-HEP-466}
\preprint{May, 2001}

\title{CP Violation in a Supersymmetric $SO(10) \times U(2)_{F}$ Model} 


\author{Mu-Chun Chen}
\email[]{mu-chun.chen@colorado.edu}
\author{K.T. Mahanthappa}
\email[]{ktm@verb.colorado.edu}
\affiliation{Department of Physics, University of Colorado, 
Boulder, CO80309-0390, U.S.A.}



\begin{abstract}
A model based on SUSY $SO(10)$ combined with $U(2)$
family symmetry constructed recently by the authors 
is generalized to include phases in the mass matrices leading to CP violation.
In contrast with the commonly used effective operator approach,
$\overline{126}$-dimensional Higgs fields are utilized to construct the Yukawa
sector. $R$-parity symmetry is thus preserved at low energies. The 
{\it symmetric} mass textures arising from the left-right symmetry breaking
chain of $SO(10)$ give rise to very good predictions for quark and lepton
masses and mixings. The prediction for $\sin 2\beta$ agrees with the average
of current bounds from BaBar and Belle. In the neutrino sector, our
predictions are in good agreement with results from atmospheric neutrino
experiments. Our model accommodates both the LOW and QVO solutions to the solar
neutrino anomaly; the matrix element for neutrinoless double beta decay is
highly suppressed. The leptonic analog of the Jarlskog invariant,
$J_{CP}^{l}$, is predicted to be of $O(10^{-2})$.  
\end{abstract}



\maketitle


$SO(10)$ has long been thought to be an attractive candidate for a
grand unified theory (GUT) for a number of reasons: First of all, it unifies
all the $15$ known fermions with the right-handed neutrino for each family
into one $16$-dimensional spinor representation. The seesaw mechanism then
arises very naturally, and the non-zero neutrino masses can thus be explained.
Since a complete quark-lepton symmetry is achieved, it has the promise for
explaining the pattern of fermion masses and mixing. Because $B-L$ contained
in $SO(10)$ is broken in symmetry breaking chain to SM, it also has the
promise for baryogenesis. Recent atmospheric neutrino oscillation data from
Super-Kamiokande indicates non-zero neutrino masses. This in turn gives very
strong support to the viability of $SO(10)$ as a GUT group. Models based on
$SO(10)$ combined with discrete or continuous family symmetry have been
constructed to understand the flavor problem~\cite{Chen:2000fp,Albright:2000sz}.
Most of the models utilize ``lopsided'' mass textures which usually require
more parameters and therefore are less constrained. Furthermore, the
right-handed neutrino Majorana mass operators in most of these models are made
out of $16_{H} \times 16_{H}$ which breaks the $R$-parity at a very
high scale. We have recently constructed a
realistic model based on supersymmetric $SO(10)$ combined with $U(2)$ family
symmetry~\cite{Chen:2000fp} (referred to as ``CM'' hereafter) which
successfully predicts the low energy fermion masses and mixings. Since we
utilize {\it symmetric} mass textures and $\overline{126}$-dimensional Higgs
representations  
for the right-handed neutrino Majorana mass operator, our
model is more constrained in addition to having $R$-parity
conserved~\cite{Mohapatra:su}. The aim of this paper is to generalize this
model to include phases in the mass matrices which lead to CP violation. We
first summarize our model followed by analytic analyses of the complex mass
textures. And then the implications of our model for neutrino mixing and CP
violation are presented.  

{\bf The Model:}
The details of our model based on $SO(10) \times U(2)_{F}$ are contained 
in CM. The following is an outline of its salient features. In order to
specify the superpotential uniquely, we invoke 
$Z_{2} \times Z_{2} \times Z_{2}$ discrete symmetry. The matter fields are
\begin{displaymath} 
\psi_{a} \sim (16,2)^{-++} \quad (a=1,2), \qquad 
\psi_{3} \sim (16,1)^{+++} 
\end{displaymath}
where $a=1,2$ and the subscripts refer to family indices; the superscripts 
$+/-$ refer to $(Z_{2})^{3}$ charges. The Higgs fields which break $SO(10)$
and give rise to mass matrices upon acquiring VEV's are
\begin{eqnarray}
(10,1):\quad & T_{1}^{+++}, \quad T_{2}^{-+-},\quad
T_{3}^{--+}, \quad T_{4}^{---}, \quad T_{5}^{+--} \nonumber\\ 
(\overline{126},1):\quad & \overline{C}^{---}, \quad \overline{C}_{1}^{+++},
\quad \overline{C}_{2}^{++-}  
\end{eqnarray}
Higgs representations $10$ and $\overline{126}$ give rise to Yukawa couplings
to the matter fields which are symmetric under the interchange of family
indices. The left-right symmetry breaking chain of $SO(10)$ is
\begin{equation}
\label{eq:SB}
\begin{array}{lll}
SO(10) & \longrightarrow & SU(4) \times SU(2)_{L} \times SU(2)_{R}\\
 & \longrightarrow & SU(3) \times SU(2)_{L} \times SU(2)_{R} \times
U(1)_{B-L}\\  
& \longrightarrow & SU(3) \times SU(2)_{L} \times U(1)_{Y} \\
& \longrightarrow & SU(3) \times U(1)_{EM} 
\end{array}
\end{equation}
The $U(2)$ family symmetry is broken in two steps \cite{Barbieri:1997ww} and
the mass hierarchy is produced using the Froggatt-Nielsen mechanism
\cite{Froggatt:1979nt}:
\begin{equation}
\label{eq:steps} 
U(2) \stackrel{\epsilon M}{\longrightarrow} 
U(1) \stackrel{\epsilon' M}{\longrightarrow}
nothing
\end{equation}
and $M$ is the UV-cutoff of the effective theory above which the family
symmetry is exact, and $\epsilon M$ and $\epsilon^{'} M$ are the VEV's
accompanying the flavon fields given by
\begin{eqnarray}
(1,2): \quad & \phi_{(1)}^{++-}, \quad \phi_{(2)}^{+-+}, \quad \Phi^{-+-}
\nonumber\\ 
(1,3): \quad & S_{(1)}^{+--}, \quad S_{(2)}^{---}, \quad
\Sigma^{++-} 
\end{eqnarray}
The various aspects of VEV's of Higgs and flavon fields are given in CM.

The superpotential for our model is
\begin{equation}
W = W_{Dirac} + W_{\nu_{RR}} + W_{flavon}
\end{equation}
\begin{eqnarray}
W_{Dirac}=\psi_{3}\psi_{3} T_{1}
 + \frac{1}{M} \psi_{3} \psi_{a}
\left(T_{2}\phi_{(1)}+T_{3}\phi_{(2)}\right)
\nonumber\\
+ \frac{1}{M} \psi_{a} \psi_{b} \left(T_{4} + \overline{C}\right) S_{(2)}
+ \frac{1}{M} \psi_{a} \psi_{b} T_{5} S_{(1)}
\nonumber\\
W_{\nu_{RR}}=\psi_{3} \psi_{3} \overline{C}_{1} 
+ \frac{1}{M} \psi_{3} \psi_{a} \Phi \overline{C}_{2}
+ \frac{1}{M} \psi_{a} \psi_{b} \Sigma \overline{C}_{2}
\end{eqnarray}
where $W_{flavon}$ is the superpotential involving only flavon fields which
give rise to their VEV's given in Eq.~(34) of CM~\cite{vev}. The mass matrices
then can be read from the superpotential to be 
\begin{eqnarray}
M_{u,\nu_{LR}} & = &
\left( \begin{array}{ccc}
0 & 0 & \left<10_{2}^{+} \right> \epsilon'\\
0 & \left<10_{4}^{+} \right> \epsilon & \left<10_{3}^{+} \right> \epsilon \\
\left<10_{2}^{+} \right> \epsilon' & \left<10_{3}^{+} \right> \epsilon &
\left<10_{1}^{+} \right>
\end{array} \right)
\nonumber\\
 & = & 
\left( \begin{array}{ccc}
0 & 0 & r_{2} \epsilon'\\
0 & r_{4} \epsilon & \epsilon \\
r_{2} \epsilon' & \epsilon & 1
\end{array} \right) M_{U}
\end{eqnarray}
\begin{eqnarray}
M_{d,e} & = & 
\left(\begin{array}{ccc}
0 & \left<10_{5}^{-} \right> \epsilon' & 0 \\
\left<10_{5}^{-} \right> \epsilon' &  (1,-3)\left<\overline{126}^{-} \right>
\epsilon & 0\\ 0 & 0 & \left<10_{1}^{-} \right>
\end{array} \right)
\nonumber\\
 & = & 
\left(\begin{array}{ccc}
0 & \epsilon' & 0 \\
\epsilon' &  (1,-3) p \epsilon & 0\\
0 & 0 & 1
\end{array} \right) M_{D}
\end{eqnarray}
where
$M_{U} \equiv \left<10_{1}^{+} \right>$, 
$M_{D} \equiv \left<10_{1}^{-} \right>$, 
$r_{2} \equiv \left<10_{2}^{+} \right> / \left<10_{1}^{+} \right>$, 
$r_{4} \equiv \left<10_{4}^{+} \right> / \left<10_{1}^{+} \right>$ and
$p \equiv \left<\overline{126}^{-}\right> / \left<10_{1}^{-} \right>$.
The right-handed neutrino mass matrix is  
\begin{eqnarray}
M_{\nu_{RR}} & = &  
\left( \begin{array}{ccc}
0 & 0 & \left<\overline{126}_{2}^{'0} \right> \delta_{1}\\
0 & \left<\overline{126}_{2}^{'0} \right> \delta_{2} 
& \left<\overline{126}_{2}^{'0} \right> \delta_{3} \\ 
\left<\overline{126}_{2}^{'0} \right> \delta_{1}
& \left<\overline{126}_{2}^{'0} \right> \delta_{3} &
\left<\overline{126}_{1}^{'0} \right> \end{array} \right)
\nonumber\\
 & = & 
\left( \begin{array}{ccc}
0 & 0 & \delta_{1}\\
0 & \delta_{2} & \delta_{3} \\ 
\delta_{1} & \delta_{3} & 1
\end{array} \right) M_{R}
\label{Mrr}
\end{eqnarray}
with $M_{R} \equiv \left<\overline{126}^{'0}_{1}\right>$.
(This is one of the five sets of symmetric texture combinations (labeled set
(v)) proposed by Ramond, Roberts and Ross~\cite{Ramond:1993kv}.) 
Here, the superscripts $+/-/0$ refer to the sign of the hypercharge. 
It is to be noted that there is a factor of $-3$ difference between the $(22)$
elements of mass matrices $M_{d}$ and $M_{e}$. This is due to the CG
coefficients associated with $\overline{126}$; as a consequence, we obtain the
phenomenologically viable Georgi-Jarlskog relations~\cite{Georgi:1979df}.

In CM, VEV's were taken to be real. In general, all VEV's are complex
leading to {\it spontaneous} CP violation which is the subject matter of this
paper. 

{\bf Analytical Analysis of Mass Textures:}
In our convention, the Yukawa couplings $Y_{i}, \; (i=u,d,e,\nu_{LR})$ are
defined in the way that the left-handed fields are on the right, and the charge
conjugate fields are on the left. In the quark sector, they read
\begin{displaymath} 
\mathcal{L}_{mass} = -Y_{u} \overline{U}_{R} Q_{L} H_{u} 
- Y_{d} \overline{D}_{R} Q_{L} H_{d} + h.c.  
\end{displaymath}
And the charged current interaction is given by
\begin{displaymath}
\mathcal{L}_{cc} = \frac{g}{\sqrt{2}}
(W_{\mu}^{+} \overline{U}_{L} \gamma_{\mu}
D_{L}) + h.c. 
\end{displaymath}
Here we have written the Lagrangian in the weak basis. 
The Yukawa matrices are diagonalized by the bi-unitary transformations
\begin{eqnarray}
Y_{u}^{diag} = V_{u_{R}} Y_{u} V_{u_{L}}^{\dagger} 
= diag(y_{u},y_{c},y_{t})
\nonumber\\
Y_{d}^{diag} = V_{d_{R}} Y_{d} V_{d_{L}}^{\dagger} 
= diag(y_{d},y_{s},y_{b})
\end{eqnarray}
where $V_{R}$ and $V_{L}$ are the right-handed and left-handed rotations
respectively, and all the eigenvalues $y_{i}$'s are real and non-negative. 
To extract the left-handed (right-handed) rotation matrices, we need to
consider the diagonalization of the hermitian quantity 
$Y^{\dagger}Y$  ($Y Y^{\dagger}$). 
The Cabbibo-Kobayashi-Maskawa (CKM) matrix is then given by 
\begin{equation}
V_{CKM} = V_{u_{L}} V_{d_{L}}^{\dagger}
= \left(
\begin{array}{ccc}
V_{ud} & V_{us} & V_{ub}\\
V_{cd} & V_{cs} & V_{cb}\\
V_{td} & V_{ts} & V_{tb}
\end{array}
\right)
\end{equation}
The unitary matrix $V_{CKM}$ has in general $6$ phases. By phase
redefinition of the quarks, one can remove $5$ of them. The remaining one
phase is one of the sources for CP violation. A parameterization
independent measure for the CP violation is the Jarlskog invariant
\cite{Jarlskog:1985cw}, defined as  
$J_{CP}^{q} \equiv Im \{ V_{11}V_{12}^{\ast}V_{21}^{\ast}V_{22} \}$. 
And the three angles of the CKM unitarity triangle are defined as
\begin{equation}
\begin{array}{ll}
\alpha \equiv 
Arg(-\frac{V_{td}V_{tb}^{\ast}}{V_{ud}V_{ub}^{\ast}}), & \quad
\beta \equiv
Arg(-\frac{V_{cd}V_{cb}^{\ast}}{V_{td}V_{tb}^{\ast}})\\
\gamma \equiv
Arg(-\frac{V_{ud}V_{ub}^{\ast}}{V_{cd}V_{cb}^{\ast}}) &
\end{array}
\end{equation}

In the lepton sector, the charged lepton Yukawa matrix is diagonalized 
by
\begin{equation}
Y_{e}^{diag} = V_{e_{R}} Y_{e} V_{e_{L}}^{\dagger} 
= diag(y_{e},y_{\mu},y_{\tau})
\end{equation}
And the effective light left-handed Majorana neutrino mass matrix 
(obtained after using the seesaw mechanism) 
is diagonalized by an unitary matrix $V_{\nu_{LL}}$:
\begin{equation}
M_{\nu_{LL}}^{diag} = V_{\nu_{LL}} M_{\nu_{LL}} V_{\nu_{LL}}^{T}
= diag(m_{\nu_{1}},m_{\nu_{2}},m_{\nu_{3}})
\end{equation}
where the eigenvalues $y_{e,\mu,\tau}$ and 
$m_{\nu_{1},\nu_{2},\nu_{3}}$ are real and non-negative. 
The leptonic mixing matrix, the Maki-Nakagawa-Sakata (MNS) matrix,
can be parameterized by
\begin{widetext}
\begin{equation}
U_{MNS} \equiv V_{e_{L}}V_{\nu_{LL}}^{\dagger}
=
\left(
\begin{array}{ccc}
c_{12}c_{13} &
s_{12}c_{13} &
s_{13}\\
-s_{12}c_{23}-c_{12}s_{23}s_{13}e^{i\delta_{l}} &
-c_{12}c_{23}-s_{12}s_{23}s_{13}e^{i\delta_{l}} &
s_{23}c_{13}e^{i\delta_{l}}\\
s_{12}s_{23}-c_{12}c_{23}s_{13}e^{i\delta_{l}} &
-c_{12}s_{23}-s_{12}c_{23}s_{13}e^{i\delta_{l}} &
c_{23}c_{13}e^{i\delta_{l}}
\end{array}
\right)
\left(
\begin{array}{ccc}
1 & &\\
& e^{i \frac{\alpha_{21}}{2}} & \\
& & e^{i\frac{\alpha_{31}}{2}}
\end{array}
\right)
\end{equation}
\end{widetext}
Note that the Majorana condition,
\begin{equation}
C(\overline{\nu}_{j})^{T} = \nu_{j}
\end{equation}
where $C$ is the charge conjugate operator, forbids the rephasing of the
Majorana fields. Therefore, we can only remove $3$ of the $6$ phases
present in the unitary matrix $U_{MNS}$ by redefing the charged lepton fields. 
Note that $U_{MNS}$ is the product of an unitary matrix, analogous to the CKM
matrix which has one phase (the so-called universal phase), $\delta_{l}$,
and a diagonal phase matrix which contains two phases 
(the so-called Majorana phases), $\alpha_{21}$ and $\alpha_{31}$.
The leptonic analog of the Jarlskog invariant, which measures the
CP violation due to the universal phase, is given by
\begin{equation}
J_{CP}^{l} \equiv Im\{ U_{\mu 2} U_{e3} U_{\mu 3}^{\ast} U_{e2}^{\ast} \}
\end{equation}
For the Majorana phases, the rephasing invariant CP violation measures are
\cite{Nieves:1987pp}
\begin{equation}
S_{1} \equiv Im\{ U_{e1} U_{e3}^{\ast} \}, \qquad
S_{2} \equiv Im\{ U_{e2} U_{e3}^{\ast} \}
\end{equation}
From $S_{1}$ and $S_{2}$, one can then determine the Majorana phases
\begin{equation}
\begin{array}{l}
\cos \alpha_{31} = 1 - 2 \frac{S_{1}^{2}}{\vert U_{e1} \vert^{2}
\vert U_{e3} \vert^{2}}\\
\cos (\alpha_{31} - \alpha_{21})
= 1 - 2 \frac{S_{2}^{2}}{\vert U_{e2} \vert^{2}
\vert U_{e3} \vert^{2}}
\end{array}
\end{equation}

The Lagrangian is invariant under the following weak-basis phase
transformations  
\begin{eqnarray}
U_{L} \rightarrow U_{L}^{'} = K_{L}^{+} U_{L}, \quad &
D_{L} \rightarrow D_{L}^{'} = K_{L}^{+} D_{L}\nonumber\\
U_{R} \rightarrow U_{R}^{'} = K_{R}^{u} U_{R}, \quad &
D_{R} \rightarrow D_{R}^{'} = K_{R}^{d} D_{R}
\end{eqnarray}
\begin{equation}
Y_{u} \rightarrow Y_{u}^{'} = K_{R}^{u} Y_{u} K_{L}, \quad
Y_{d} \rightarrow Y_{d}^{'} = K_{R}^{d} Y_{u} K_{L}
\end{equation}
where $K_{L}^{+}$, $K_{R}^{u}$, and $K_{R}^{d}$ are diagonal phase matrices.
We are interested in complex symmetric textures resulting from the
$SO(10)$ relations: 
\begin{eqnarray}
Y_{u} & = & \left(
\begin{array}{ccc}
0 & 0 & a e^{i\gamma_{a}} \\
0 & b e^{i\gamma_{b}} & c e^{i\gamma_{c}} \\
a e^{i\gamma_{a}} & c e^{i\gamma_{c}} & e^{i\gamma}
\end{array}
\right) d
\nonumber\\
Y_{d} & = & \left(
\begin{array}{ccc}
0 & e e^{i\gamma_{e}} & 0 \\
e e^{i\gamma_{e}} & f e^{i\gamma_{f}} & 0 \\
0 & 0 & e^{i\gamma_h}
\end{array}
\right) h
\nonumber\\
Y_{\nu_{LR}} & = & \left(
\begin{array}{ccc}
0 & 0 & a e^{i\gamma_{a}} \\
0 & b e^{i\gamma_{b}} & c e^{i\gamma_{c}} \\
a e^{i\gamma_{a}} & c e^{i\gamma_{c}} & e^{i\gamma}
\end{array}
\right) d
\nonumber\\
Y_{e} & = & \left(
\begin{array}{ccc}
0 & e e^{i\gamma_{e}} & 0 \\
e e^{i\gamma_{e}} & -3 f e^{i\gamma_{f}} & 0 \\
0 & 0 & e^{i\gamma_h}
\end{array}
\right) h\nonumber
\end{eqnarray}
with $a \simeq b \ll c \ll 1$ and $e \ll f \ll 1$. 
The weak-basis phase transformations mentioned above enable us to reduce the
number of phases to two, $\theta \equiv (\gamma_{b}-2\gamma_{c}-\gamma)$ and 
$\xi \equiv (\gamma_{e}-\gamma_{f}+\gamma_{c}-\gamma_{a})$. 
Then we have 
\begin{eqnarray}
Y_{u, \nu_{LR}} & = & \left(
\begin{array}{ccc}
0 & 0 & a\\
0 & b e^{i\theta} & c\\
a & c & 1
\end{array}
\right) d
\nonumber\\ 
Y_{d,e} & = & \left(
\begin{array}{ccc}
0 & e e^{-i\xi} & 0 \\
e e^{i\xi} & (1,-3) f & 0 \\
0 & 0 & 1
\end{array}
\right) h
\label{phaseremoved}
\end{eqnarray}
We then diagonalize the mass matrices analytically. In the
leading order, the mass eigenvalues are given by
\begin{eqnarray}
\label{eq:masses}
m_{u} & \simeq & \frac{a^{2} b d}{\vert -b e^{i\theta} + c^{2} \vert} v_{u}
\nonumber\\
m_{c} & \simeq & \vert (-b e^{i\theta}+c^{2})\vert d v_{u}
\nonumber\\
m_{t} & \simeq & d v_{u} 
\nonumber\\
m_{d} & \simeq & \frac{e^{2} f h}{e^{2}+f^{2}}v_{d}
\nonumber\\
m_{s} & \simeq & \frac{(2e^{2} f + f^{3}) h}{e^{2} + f^{2}}v_{d}
\nonumber\\
m_{b} & = & hv_{d}
\nonumber\\
m_{e} & \simeq & \frac{3e^{2} f h}{e^{2} + 9 f^{2}}v_{d}
\nonumber\\
m_{\mu} & \simeq & \frac{6e^{2} f h + 27 f^{3} h}{e^{2} + 9 f^{2}}v_{d}
\nonumber\\
m_{\tau} & = & hv_{d}
\end{eqnarray}
The CKM matrix elements are given by 
\begin{eqnarray}
\label{ckm}
V_{ud} & \simeq & (1-\frac{1}{2}\frac{e^{2}}{f^{2}}) 
- (\frac{e}{f})(\frac{a}{c})e^{-i\delta_{q}} 
\nonumber\\
V_{us} & \simeq & \frac{e}{f} + \frac{a}{c} e^{-i\delta_{q}}
\nonumber\\
V_{ub} & \simeq & a e^{-i\delta_{q}}
\nonumber\\
V_{cd} & \simeq &  - \frac{a}{c}e^{i\delta_{q}} - \frac{e}{f}
\nonumber\\
V_{cs} & \simeq & 
(1-\frac{1}{2}\frac{e^{2}}{f^{2}}) 
- (\frac{e}{f})(\frac{a}{c})e^{i\delta_{q}} 
\nonumber\\
V_{cb} & \simeq &  c 
\nonumber\\
V_{td} & \simeq & \frac{e}{f} c
\nonumber\\
V_{ts} & \simeq &  -c 
\nonumber\\
V_{td} & \simeq &  1 - \frac{1}{2} c^{2} 
\end{eqnarray}
The phase $\delta_{q}$ is given by
\begin{eqnarray}
\delta_{q} & \simeq & \tan^{-1} (-\frac{a}{c} \sin (\theta^{'}-\xi))
\nonumber\\ 
 & - & \tan^{-1} (-\frac{e}{f} \frac{a}{c} \sin(\theta^{'}-\xi))
 + (\theta^{'}-\xi)
\end{eqnarray}
where
\begin{equation}
\label{eq:phase}
\theta^{'} \simeq \tan^{-1} (
b\sin\theta - 2 \frac{b^{3}}{a^{2}}\sin^{2}\theta
-2 \frac{b}{a^{2}} \sin \theta
)
\end{equation} 
The form of the CKM matrix in Eq.~(\ref{ckm}) is also the one favored in   
\cite{Rasin:1997pn}.

In the lepton sector, similar diagonalization is carried out, and the charged
lepton diagonalization matrix is
\begin{equation}
V_{e} = V_{d} \; (b \rightarrow -3b)
\end{equation}
We obtain bimaximal mixing in the neutrino sector and 
$\Delta m_{atm}^{2} \gg \Delta m_{\odot}^{2}$ by choosing
\cite{Chen:2000fp}
\begin{equation}
M_{\nu_{LL}} \simeq
\left(
\begin{array}{ccc}
0 & 0 & t\\
0 & 1 & 1\\
t & 1 & 1
\end{array}
\right)
\frac{d^{2}v_{u}^{2}}{M_{R}}
\end{equation}
In the basis where the Yukawa matrices take the forms of
Eq.~(\ref{phaseremoved}), this implies that the right-handed neutrino mass
matrix, Eq.~(\ref{Mrr}), has the following elements: 
\begin{eqnarray}
\label{delta}
\delta_{1} & \simeq & \frac{a^{2}}{2a-2ac+c^{2}t}
\nonumber\\
\delta_{2} & \simeq &
\frac{b^{2}t e^{2i\theta}}{2a-2ac+c^{2}t}
\nonumber\\
\delta_{3} & \simeq &
\frac{a(c-be^{i\theta})+bct e^{i\theta}} {2a-2ac+c^{2}t}
\end{eqnarray}
The three eigenvalues of $M_{\nu_{LL}}$ are
\begin{eqnarray}
m_{\nu_{1}} & \simeq & (\frac{t}{\sqrt{2}}-\frac{t^{2}}{8})
\frac{d^{2}v_{u}^{2}}{M_{R}}
\nonumber\\
m_{\nu_{2}} & \simeq & (\frac{t}{\sqrt{2}}+\frac{t^{2}}{8})
\frac{d^{2}v_{u}^{2}}{M_{R}}
\nonumber\\
m_{\nu_{3}} & \simeq & (2 + \frac{t^{2}}{4})
\frac{d^{2}v_{u}^{2}}{M_{R}}
\end{eqnarray}
Note that since $V_{e_{L}}$ is approximately an identity matrix, 
$U_{MNS} \approx V_{\nu_{LL}}^{\dagger}$. 
Consequently, with the present experimental status, it is not possible to make
unique predictions for the leptonic CP violating phases from the fitting of the
absolute values of the $MNS$ matrix elements, as was done in the quark sector.
We therefore assume that the light neutrino mass matrix to be real. And the
leptonic CP violation will solely due to the phase present in $Y_{e}$.

{\bf RGE Analysis:}
We use the following inputs at $M_{Z}=91.187 \; GeV$:
\begin{eqnarray}
m_{u} & = & 2.32 \; MeV (2.33^{+0.42}_{-0.45}) \nonumber\\ 
m_{c} & = & 677 \; MeV (677^{+56}_{-61}) \nonumber\\ 
m_{t} & = & 182 \; GeV (181^{+}_{-}13) \nonumber\\
m_{e} & = & 0.485 \; MeV (0.486847) \nonumber\\
m_{\mu} & = & 103 \; MeV (102.75) \nonumber\\ 
m_{\tau} & = & 1.744 \; GeV (1.7467) \nonumber\\
\vert V_{us} \vert & = & 0.222 (0.219-0.224) \nonumber\\
\vert V_{ub} \vert & = & 0.0039 (0.002-0.005) \nonumber\\
\vert V_{cb} \vert & = & 0.036 (0.036-0.046) \nonumber
\end{eqnarray} 
where the values extrapolated from experimental data are given inside the
parentheses~\cite{Fusaoka:1998vc}. 
These values correspond to the following set of input parameters
at the GUT scale,  $M_{GUT} = 1.03 \times 10^{16} \; GeV$: 
\begin{displaymath} 
a=0.00246, \quad b=3.50 \times 10^{-3}, \quad  c=0.0320
\end{displaymath}
\begin{displaymath} 
\quad d=0.650, \quad \theta=0.110
\end{displaymath} 
\begin{displaymath}
e=4.03 \times 10^{-3}, \quad f=0.0195
\end{displaymath}
\begin{equation}
\label{input}
h=0.0686, \quad \xi=-0.720
\end{equation}
\begin{equation}
g_{1}=g_{2}=g_{3}=0.746
\end{equation}
the one-loop renormalization group equations for the MSSM spectrum with three
right-handed neutrinos 
\cite{Babu:1993qv}
are solved numerically down to the effective
right-handed neutrino mass scale, $M_{R}$. At $M_{R}$, the seesaw mechanism  
is implemented. We then solve the two-loop RGE's for the MSSM spectrum 
\cite{Arason:1992ic}
down to the SUSY breaking scale, taken to be $m_{t}(m_{t})=176.4 \; GeV$, and
then the SM RGE's from $m_{t}(m_{t})$ to the weak scale, $M_{Z}$. 
We assume that 
$\tan \beta \equiv \frac{v_{u}}{v_{d}} = 10$,  with 
$v_{u}^{2} + v_{d}^{2} = (246/\sqrt{2} \; GeV) ^{2}$. At the weak scale
$M_{Z}$, the predictions for 
$\alpha_{i}
\equiv \frac{g_{i}^{2}}{4\pi}$ are   
\begin{displaymath} 
\alpha_{1}=0.01663,
\quad \alpha_{2}=0.03374, 
\quad \alpha_{3}=0.1242 
\end{displaymath}
These values compare very well with the values extrapolated to $M_{Z}$ from the
experimental data~\cite{Fusaoka:1998vc}, 
$(\alpha_{1},\alpha_{2},\alpha_{3})=
(0.01696,0.03371,0.1214 {\scriptstyle + \atop -} 0.0031)$.
The predictions at the weak scale $M_{Z}$ for the
charged fermion masses, the CKM matrix elements and the CP violation measures, 
are summarized in TABLE~\ref{table:ckm} (after taking
into account the SUSY threshold correction
\cite{Hall:1994gn}, 
$\Delta_{b} = -0.15$)
along with the experimental values extrapolated to $M_{Z}$ 
\cite{Fusaoka:1998vc}.

\begin{table}
\caption{
The predictions for the charged fermion masses, the 
CKM matrix elements and the CP violation measures.
\label{table:ckm}}
\begin{ruledtabular}
\begin{tabular}{l c | c c l c c c l}
 & & & & experimental results \qquad \qquad 
 & & & & predictions at $M_{z}$ \\ 
& & & & extrapolated to $M_{Z}$ \qquad \qquad
 & & & & \\ 
\hline
$\frac{m_{s}}{m_{d}}$  
& & & & $17 \sim 25$  
& & & & $25$\\
$m_{s}$ 
& & & & $93.4^{+11.8}_{-13.0}MeV$  
& & & & $85.66 MeV$\\
$m_{b}$ 
& & & & $3.00^{+}_{-}0.11GeV$  
& & & & $3.147 GeV$\\
\hline
$\vert V_{ud} \vert$
& & & & $0.9745-0.9757$ 
& & & & $0.9751$\\
$\vert V_{cd} \vert$  
& & & & $0.218-0.224$  
& & & & $0.2218$\\
$\vert V_{cs} \vert$ 
& & & & $0.9736-0.9750$  
& & & & $0.9744$\\
$\vert V_{td} \vert$  
& & & & $0.004-0.014$  
& & & & $0.005358$\\
$\vert V_{ts} \vert$  
& & & & $0.034-0.046$ 
& & & & $0.03611$\\
$\vert V_{tb} \vert$  
& & & & $0.9989-0.9993$  
& & & & $0.9993$ \\
$J_{CP}^{q}$  
& & & & $(2.71^{+}_{-}1.12) \times 10^{-5}$  
& & & & $1.748 \times 10^{-5}$ \\
$\sin 2\alpha$  
& & & & $-0.95 \; - \; 0.33$
& & & & -0.8913\\
$\sin 2\beta$ 
& & & & $0.59^{+}_{-}0.14^{+}_{-}0.05$ (BaBar)
& & & & $0.7416$ \\
& & & & $0.99^{+}_{-}0.14^{+}_{-}0.06$(Belle)
& & & &\\
$\gamma$  
& & & & $34^{0}-82^{0}$
& & & & $34.55^{0} \; (0.6030 rad)$
\end{tabular}
\end{ruledtabular}
\end{table}

The current result from the atmospheric neutrino experiment is
\cite{Fukuda:2000np}
\begin{eqnarray}
\Delta m_{23}^{2} & = & (1.3 - 8) \times 10^{-3} eV^{2}
(\text{best fit}: 3.3 \times 10^{-3}) \nonumber\\
\sin^{2} 2\theta & = & 1 \nonumber
\end{eqnarray}
In the solar neutrino sector, 
the current best fit from the solar neutrino experiment data for each of the
solution is given by
\cite{Gonzalez-Garcia:2000ae}:
\pagebreak
\begin{eqnarray}
LMA: & \mbox{(large angle MSW with larger mass squared)}
\nonumber\\
& \Delta m^{2} = 3.2
\times 10^{-5} eV^{2} \nonumber\\
& \tan^2 \theta=0.33,
\quad \sin^{2}2\theta=0.75
\nonumber\\
LOW: & \mbox{(large angle MSW with smaller mass squared)}
\nonumber\\
& \Delta m^{2} = 1 \times 10^{-7} eV^{2} \nonumber\\  
& \tan^{2} \theta = 0.67, \quad \sin^{2} 2\theta=0.96
\nonumber\\
SMA: & \mbox{(small angle MSW)}
\nonumber\\
& \Delta m^{2} = 5.0 \times 10^{-6} eV^{2} \nonumber\\
& \tan^{2} \theta = 0.0006, \quad \sin^{2}2\theta=0.0024
\nonumber\\
QVO: & \mbox{(quasi-vacuum oscillation)}
\nonumber\\
& \Delta m^{2} = 8.6 \times 10^{-10} eV^{2} \nonumber\\
& \tan^{2} \theta = 1.5, \quad \sin^{2}2\theta=0.96
\nonumber
\end{eqnarray}
At present, none of these solutions can be ruled out \cite{Bahcall:2001hv}. 
The current bound on $\vert U_{e\nu_{3}} \vert$ from CHOOZ experiment is  
\cite{Apollonio:1999ae}
\begin{displaymath}
\vert U_{e\nu_{3}} \vert \le 0.16
\end{displaymath}
Our model favors both the LOW and QVO solution. 
The LOW solution is obtained with the following set of
parameters:  
\begin{equation}
M_{R} = 2.012 \times 10^{13} GeV, \quad t = 0.088
\end{equation}
which give rise to, using Eq.~(\ref{delta}),
\begin{eqnarray}
\delta_{1} & = & 0.001247 \nonumber\\ 
\delta_{2} & = & 2.221 \times 10^{-4} e^{i \; (0.2201)} \nonumber\\ 
\delta_{3} & = & 0.01648 e^{-i \;(0.001711)} \nonumber
\end{eqnarray}
This gives rise to three light neutrino masses
\begin{eqnarray}
m_{\nu_{1}} & = & 0.001711 \; eV \nonumber\\ 
m_{\nu_{2}} & = & 0.001762 \; eV \nonumber\\
m_{\nu_{3}} & = & 0.05438 \; eV \nonumber
\end{eqnarray}
and the squared-mass differences are 
\begin{eqnarray}
\Delta m_{atm}^{2} & = & 2.954 \times 10^{-3} \; eV^{2} \nonumber\\ 
\Delta m_{\odot}^{2} & = & 1.769 \times 10^{-7} \; eV^{2}
\end{eqnarray}
And the MNS matrix is given by
\begin{displaymath}
\vert U_{MNS} \vert = 
\left(
\begin{array}{ccc}
0.6743 & 0.7346 & 0.07497\\
0.5427 & 0.4322 & 0.7202\\
0.5008 & 0.5230 & 0.6897
\end{array}
\right)
\end{displaymath}
This translates into 
\begin{eqnarray}
\sin^{2} 2 \theta_{atm} \equiv 4 \vert U_{\mu \nu_{3}} \vert^{2}
(1-\vert U_{\mu \nu_{3}} \vert^{2}) = 0.9986
\nonumber\\
\sin^{2} 2 \theta_{\odot} \equiv 4 \vert U_{e \nu_{2}} \vert^{2}
(1-\vert U_{e \nu_{2}} \vert^{2}) = 0.9937
\nonumber
\end{eqnarray}
and the MNS matrix element $\vert U_{e\nu_{3}} \vert$ is predicted to be
$0.07497$, in very good agreement with the experimental result. The leptonic
analog of the Jarlskog invariant is predicted to be 
\begin{displaymath}
J_{CP}^{l} \equiv Im \{ U_{11} 
U_{12}^{\ast} 
U_{21}^{\ast} U_{22} \}
= -0.008147
\end{displaymath}
The matrix element for the neutrinoless double beta  
$(\beta\beta_{0\nu})$ decay, 
$\vert < m > \vert$, 
is given in terms of the rephasing invariant quantities by
\begin{eqnarray}
\vert < m > \vert ^{2} & = &
m_{1}^{2} \vert U_{e1} \vert^{4} +  m_{2}^{2} \vert U_{e2} \vert^{4}
+ m_{3}^{2} \vert U_{e3} \vert^{4} 
\nonumber\\
& + & 
2m_{1}m_{2} \vert U_{e1} \vert^{2} \vert U_{e2} \vert^{2} \cos\alpha_{21} 
\nonumber\\
& + & 
2m_{1}m_{3} \vert U_{e1} \vert^{2} \vert U_{e3} \vert^{2} \cos\alpha_{31} 
\nonumber\\
& + & 
2m_{2}m_{3} \vert U_{e2} \vert^{2} \vert U_{e3} \vert^{2} 
\cos(\alpha_{31}-\alpha_{21})\nonumber
\\
\end{eqnarray}
The current upper bound for $\vert < m > \vert$ 
from the experiment is $0.2 \; eV$  \cite{Baudis:1999xd}. 
The two Majorana phases $(\alpha_{31},\alpha_{21})$ are $(0.9708,-1.326)$; 
they give rise to a highly suppressed 
$\vert<m>\vert = 1.359 \times 10^{-3} \; eV$\cite{Chen:2001gk}. 
The three heavy neutrino masses are given by 
\begin{eqnarray}
\label{Mr:low}
M_{1} & = & 9.412 \times 10^{7} GeV \nonumber\\
M_{2} & = & 1.486 \times 10^{9} GeV \nonumber\\
M_{3} & = & 2.013 \times 10^{13} GeV \nonumber
\end{eqnarray}

The QVO solution is obtained with:
\begin{equation}
M_{R} = 1.217 \times 10^{14} GeV, \quad t = 0.0142
\end{equation}
which give rise to, 
\begin{eqnarray}
\delta_{1} & = & 0.001267 \nonumber\\
\delta_{2} & = & 3.641 \times 10^{-5} e^{i \; (0.2201)} \nonumber\\ 
\delta_{3} & = & 0.01502 e^{-i \;(0.01074)} \nonumber
\end{eqnarray}
The three light neutrino masses are predicted to be
\begin{eqnarray}
m_{\nu_{1}} & = &  2.856 \times 10^{-4} \; eV \nonumber\\
m_{\nu_{2}} & = &  2.870 \times 10^{-4} \; eV \nonumber\\
m_{\nu_{3}} & = &  5.587 \times 10^{-2} \; eV \nonumber
\end{eqnarray}
and the squared-mass differences are 
\begin{eqnarray}
\Delta m_{atm}^{2} & = & 3.122 \times 10^{-3} \; eV^{2} \nonumber\\ 
\Delta m_{\odot}^{2} & = & 7.584 \times 10^{-10} \; eV^{2} \nonumber
\end{eqnarray}
And the MNS matrix is given by
\begin{displaymath}
\vert U_{MNS} \vert = 
\left(
\begin{array}{ccc}
0.6794 & 0.7318 & 0.05310\\
0.5302 & 0.4513 & 0.7178\\
0.5072 & 0.5107 & 0.6942
\end{array}
\right)
\end{displaymath}
This translates into 
\begin{eqnarray}
\sin^{2} 2 \theta_{atm} = 0.9991, \quad
\sin^{2} 2 \theta_{\odot} = 0.9950
\nonumber
\end{eqnarray}
and the MNS matrix element $\vert U_{e\nu_{3}} \vert$ is predicted to be
$0.05310$, in very good agreement with the experimental result. The leptonic
analog of the Jarlskog invariant is predicted to be 
\begin{displaymath}
J_{CP}^{l} = -0.008110
\end{displaymath}
The two Majorana phases $(\alpha_{31},\alpha_{21})$ are $(1.385,-0.4976)$
and the matrix element for $\beta\beta_{0\nu}$ is predicted to be 
\protect\linebreak 
$\vert<m>\vert=3.07 \times 10^{-4}$ eV. 
The three heavy right-handed Majorana neutrino masses are 
\begin{eqnarray}
M_{1} & = & 3.699 \times 10^{7} GeV \nonumber\\
M_{2} & = & 2.341 \times 10^{10} GeV \nonumber\\
M_{3} & = &  1.218 \times 10^{14} GeV \nonumber
\end{eqnarray}
Our model can also produce the LMA solution by properly choosing the
values for $M_{R}$ and $t$; however, the prediction for $U_{e\nu_{3}}$
violates the experimental bound, leading to the elimination of the LMA
solution in our model. 

A few words concerning baryonic asymmetry are in order. Even though the
sphaleron effects destroy baryonic asymmetry, it could be produced as an
asymmetry in the generation of $(B-L)$ at a high scale because of lepton
number violation due to the decay of heavy right-handed Majorana 
neutrinos~\cite{Fukugita:1986hr}, 
which in turn is converted into baryonic asymmetry due to sphalerons.
But in our model this mechanism produces baryonic asymmetry of $O(10^{-13})$ 
which is too small to account for the observed value of 
$(1.7-8.3) \times 10^{-11}$~\cite{Olive:1999ij},
reasons being that the mass of the lightest right-handed Majorana neutrino
is too small {\it and} the $1-3$ family mixing of right-handed neutrinos is too
large, leading, in essence, to the violation of the out-of-equilibrium
condition required by Sakharov~\cite{Sakharov:1967dj}. 
So a mechanism other than leptogenesis is
required to explain baryonic asymmetry.

{\bf Summary:}
We have generalized our recently constructed model based on SUSY
$SO(10)$ combined with $U(2)$ family symmetry 
to include phases in the mass matrices leading to CP violation. In
contrast with the commonly used effective operator approach,
$\overline{126}$-dimensional Higgs fields are utilized to construct the Yukawa
sector. $R$-parity symmetry is thus preserved at low energies. The {\it
symmetric} mass textures arising from the left-right symmetry breaking chain
of $SO(10)$ give rise to very good predictions for quark and lepton masses and
mixings. The prediction for $\sin 2\beta$ agrees with the average of current bounds from
BaBar and Belle. In the neutrino sector, our predictions are in good agreement
with results from atmospheric neutrino experiments. Our model allows  
both the LOW and QVO solutions to the solar neutrino anomaly, and the matrix
element for $\beta \beta_{0\nu}$ decay is highly suppressed. The leptonic
analog of the Jarlskog invariant, $J_{CP}^{l}$, is predicted to be of
$O(10^{-2})$. It is interesting to note that, in the Yukawa sector, the model
predicts $(12+3)$ masses, $(6+3)$ mixing angles and $(4+3)$ phases (- the
additional $3$'s in the parentheses refer to the right-handed neutrino sector)
in terms of nine parameters given in Eq.~(\ref{input}) and $t$ and $M_{R}$, a
total of eleven parameters. 

\begin{acknowledgments}
We would like to thank Rabi Mohapatra for bringing to our attention the
interesting aspect of baryogenesis. This work was supported, in part, 
by the U.S. Department of Energy under Grant No. DE FG03-05ER40892.
\end{acknowledgments}

\bibliography{cp}

\end{document}